\documentstyle[epsf]{article}
\begin{document}
\title{\Large \bf Predicting Optimal Lengths of Random Knots }
\maketitle
\begin{center}
Akos Dobay$^{1}$, Pierre-Edouard Sottas$^{1,2}$, Jacques Dubochet$^{1}$ and \\Andrzej Stasiak$^{1}$
\vspace{4mm}
\\{\small \it $^{1}$Laboratory of Ultrastructural Analysis, University of Lausanne, 1015 Lausanne, Switzerland}
\vspace{1mm}
\\{\small \it $^{2}$Center for Neuromimetic Systems, Swiss Federal Institute of Technology, EPFL-DI, 1015 Lausanne, Switzerland}
\begin{abstract}
In thermally fluctuating long linear polymeric chain in solution, the ends come from time to time into a direct contact or a close
vicinity of each other. At such an instance, the chain can be regarded as a closed one and thus will form a knot or rather a virtual knot.
Several earlier studies of random knotting demonstrated that simpler knots
show their highest occurrence for 
shorter random walks than more complex knots.
However up to now there were no rules that could be used to predict the optimal length of a random walk, i.e. the length for which a given knot reaches its highest 
occurrence. Using numerical simulations, we show here that a power law accurately describes the relation between the optimal lengths of
random walks leading to the formation of different knots and the previously characterized lengths of ideal knots of the corresponding
type.
\end{abstract}
\vspace{2mm}
{\bf keywords:} knots, polymers, scaling laws, DNA, random walks.
\vspace{2mm}
\end{center}

\begin{figure}[htbp]
\hbox to \hsize{\hfill\epsfxsize=4.2 in \epsfbox{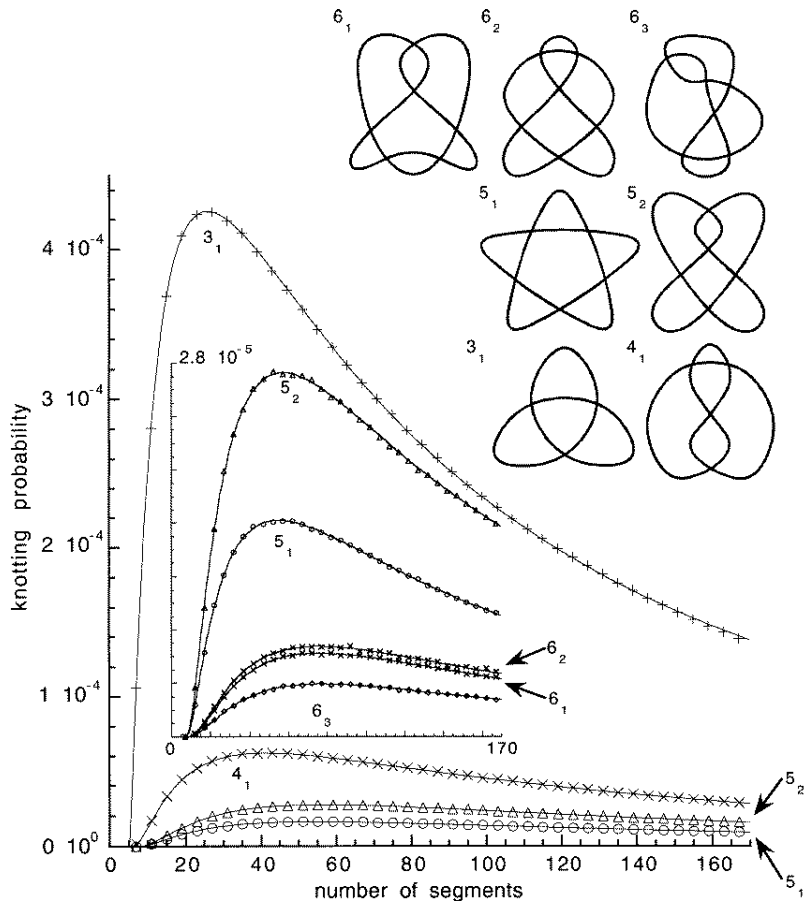}\hfill}
\caption{Probability of forming a given knot amoung all random walks of a given size is plotted as a function of the 
number of segments in the walk. Note the change of the scale 
between the main panel and the insert. Diagrams of the 
corresponding knots are drawn to visualize the differences between 
analyzed types of knots. The notations accompanying the drawn diagrams 
correspond to those in standard tables of knots \cite{21}, where the main 
number indicates the minimal number of crossings possible for this knot 
type and the index indicates the tabular position amongst the knots with 
the same minimal crossings number. Formed knot types were recognised by computing their Alexander polynomial. Since Alexander polynomial does not distinguish between left-handed and right-handed knots of the same type,
we have to group them together and therefore the drawn diagrams of the knots do not show the handedness. This polynomial has sometimes the 
same value for different knots like for example knot $6_{1}$ and $9_{46}$ \cite{22}. 
However within groups of knots with the same Alexander polynomial more 
complicated knots have such a low occurrence that their effect on the 
position of the maximum of the simplest knot within the group can be 
neglected.}
\label{fig1}
\end{figure}

\begin{figure}[htbp]
\hbox to \hsize{\hfill\epsfxsize=4.5 in \epsfbox{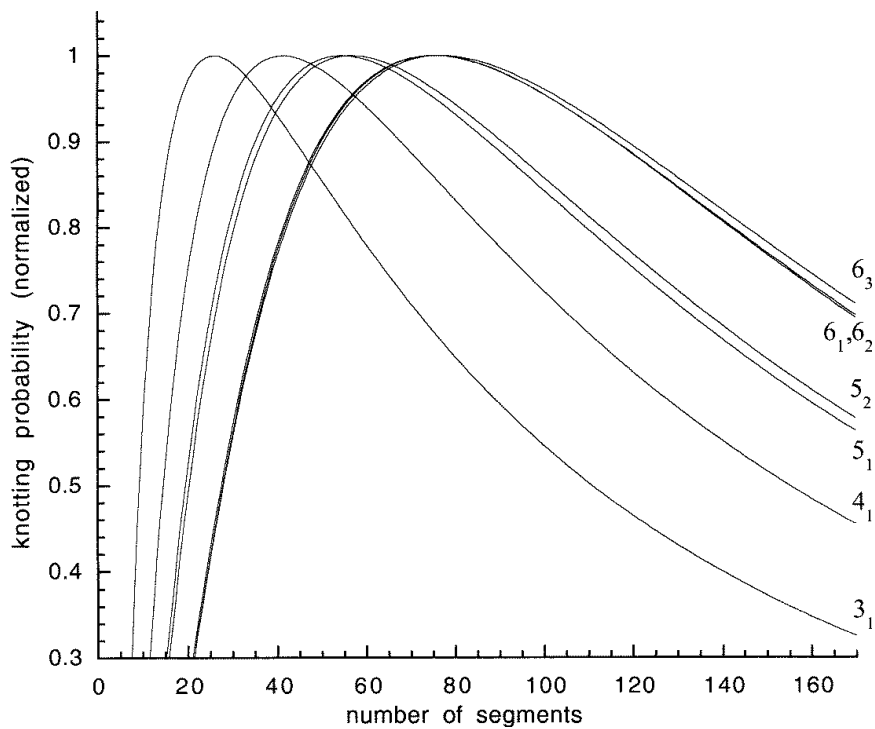}\hfill}
\caption{Normalized probability profiles for the analyzed knots.}
\label{fig2}
\end{figure}

\begin{figure}[htbp]
\hbox to \hsize{\hfill\epsfxsize=4.5 in \epsfbox{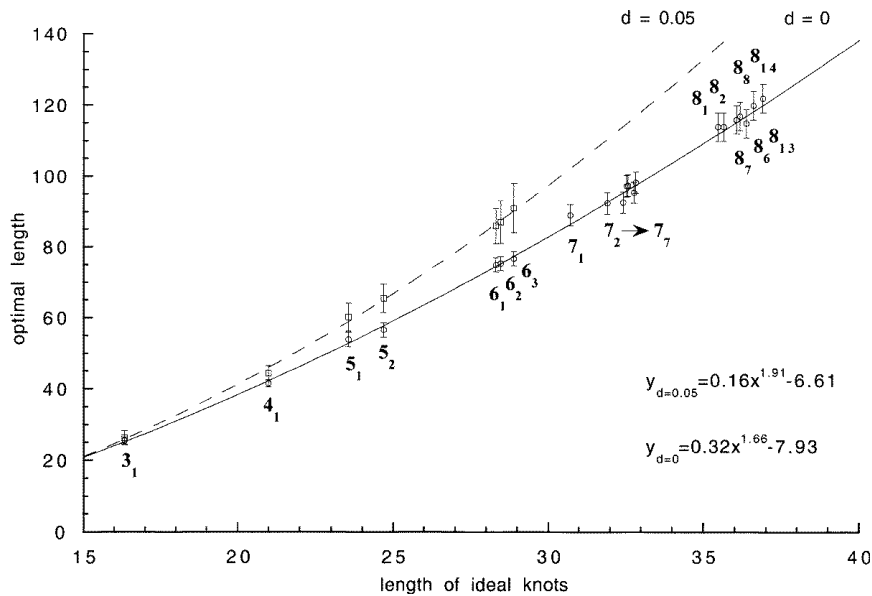}\hfill}
\caption{Relation between the length of the ideal 
geometric representations of knots \cite{23} and positions of maximal occurrence 
for the corresponding random knots. The lower curve: the optimal length of random knots with an effective diameter set to zero. The simulation data for the knots 
with up to seven crossings were fitted with a power law 
function and the best fit curve was extrapolated. Data points 
for eight crossing knots for which we obtained good statistics 
coincide with the extrapolated curve. The upper curve: data points of maximal occurrence of knots for random chains with an effective
diameter set to 0.05 of the segment length. In both cases a power law function adequately describes the relation between the optimal length of random knots and
the length of ideal knots of a given type. Best fit parameters for both cases are indicated.}
\label{fig3}
\end{figure}

A random walk can frequently lead to the formation of knots and it was proven that as the walk becomes very long the
probability of forming nontrivial  knots upon closure of such a walk tends to one \cite{1,2}. Many different
simulation approaches were used to study random knotting \cite{3,4,5,6,7}. Probably  the
most fundamental one is by simulation of ideal random chains where  each segment of the chain is of the same
length and has no thickness \cite{4,8}. In ideal random chains the neighboring segments are not correlated 
with each other and thus show the average deflection angle of 90$^{\circ}$. Ideal random chain behavior is interesting
from physical point of view as it  reflects statistical behavior of long polymer chains in so-called melt 
phase and in $\theta$ solvents where excluded volume effect vanishes \cite{8}. Highly diluted polymer chains in $\theta$ solvents are unlikely to interact with each other and therefore upon circularization will form mainly
knots rather than links. In thermally fluctuating long linear polymers the ends of the same chain can come from time to time into a close  vicinity of each other.
This can lead to a cyclization of  the polymer whereby the end closure frequently traps a
nontrivial knot on  the chain. By studying knotting in simulated ideal random chains we thus can gain insight into knotting of real
polymer chains in $\theta$ solvents and in the dense melt phase frequently used for the  preparation of such synthetic
polymeric materials like fabrics, paints or  adhesives \cite{9}. However, ideal chains do not reflect the behavior of real polymer chains in good
solvent. Intramolecular interactions cannot be neglected in these conditions, but can be well approximated by introducing an effective
diameter. When such a constraint is introduced into simulated chains one can also
model knotting of polymers in good solvents like for example knotting of DNA molecules in typical reaction buffers used for biochemical
experiments \cite{4}. Our simulations can be adjusted to both situations and we shall present here results for random chains with and
without an effective diameter.
\vspace{1.5mm}
\\Several earlier studies of random knotting showed that
simpler knots reach a maximum of their occurrence for shorter length of random  walks than this required for the formation of more complex
knots \cite{5,6,10}. In considering the equilibrium ensemble of closed walks, these studies showed that the relative frequency of occurrence of each type of knot first increases with the length of the
chain, then passes through a maximum and finally decreases exponentially at very long chains. However, these earlier studies did not
attempt to establish a relation  between the type of a knot and the optimal length of a random walk  leading to the maximal occurrence
of this knot. If we consider a thermally fluctuating polymer with ends that can stick to each other with the energy much smaller than $kT$, then from time to time these ends will stay in contact for a short period and at
this moment the polymer will form a trivial or nontrivial knot. In this study, we characterize statistical ensembles of fluctuating linear
polymers in order to find specific lengths (expressed in number of statistical segments) at which a given type of knot or rather a
virtual knot reaches its highest occurrence.
\vspace{1.5mm}
\\Recently we have  characterized
ideal geometric configurations of knots corresponding to the shortest trajectories of flexible  cylindrical tube with uniform diameter to
form a given knot \cite{11}. The  ratio of the length to diameter of the tube forming ideal configuration  of
a given knot is a topological invariant and we call it here the length of ideal  knots. Ideal knots turned out to be good
predictors of statistical  behavior of random knots. So for example the writhe of ideal  configuration of a
given knot was equal to the average writhe of thermally fluctuating polymer forming a given random
knot \cite{11}. We showed  also that electrophoretic migrations of various types of knotted DNA  molecules of
the same molecular weight or their expected sedimentation  constants were practically proportional to the length of
the corresponding  ideal knots \cite{12,13}. Therefore we decided here to check whether the length of ideal  knots is
related to the length of ideal random chains for which different knots reach their highest occurrence. To this
aim we used the following  simulation procedure. 2$\cdot$10$^{9}$ random walks of 170 segments were started and  each time
the growing end approached the starting end to a distance  smaller than the length of one segment the
configuration was saved upon  which the walk was continued for the remaining number of steps. Each  vector (segment) of the chain was randomly chosen from uniformly 
distributed vectors pointing from the center to the surface of the unit  sphere. Thus some  of the
random walks showed one or more approaches of the growing and  starting ends and we collected
2$\cdot$10$^{9}$ random walks for every number of  segments between 5 and 170. Each saved configuration with nearby ends was  then closed with
a connecting segment and the type of the formed knot was  determined by the calculation of its Alexander
polynomial \cite{7,14,15,16}.
\vspace{1.5mm}
\\For random linear walks to efficiently form different knots a compromise has to be met between the length
optimizing their close approach and the length which is sufficient to form a knot of a given type. The present analysis differs from
earlier studies \cite{4,5,10} where the statistics was based only on equilibrium knotting of closed walks. In our case, we consider
the formation of knots through the approach of the terminal segments of linear chains. Therefore not only closed chains, but also linear chains are taken into
account in our statistics.\\Figure \ref{fig1} shows the  occurrence profiles of different knots with up
to six crossings as a  function of the length of random walk which leads to the formation of these  knots. It is
visible that trefoil knots show their highest occurrence for $25\pm1$ segments while $4_{1}$ knots form most frequently for $42\pm1$
segments. The  formation of more complicated knots happens much less frequently than  this of simpler knots,
therefore in the insert in Figure \ref{fig1} a change of  scale is applied to better visualize the occurrence of more
complicated  knots. We observed that the obtained probabilities values for different knots can be well fitted with the
function
\begin{equation}
P_k(N)=a(N-N_{0})^{b}\exp(\frac{-N^{c}}{d})
\end{equation}
where for each knot $a$, $b$ and $d$ are free parameters, $c$ is an empirical constant equal
to $0.18$, $N_{0}$ is the minimal number of segments required to form a given type of knot \cite{17} without the closing segment and $N$
is the number of segments in the walk. Our fitting function was adapted from Katritch et {\it al.\/} 2000 \cite{18} but modified to take into account the probability
of cyclization. Table \ref{tableI}  lists the positions of maximal occurrence for the analyzed types of 
knots. In order to concentrate on the position of the maximum for  different knots and not on their
actual probability values we decided to  present probability profiles for each knot upon normalizing them by 
assigning a value 1 to the respective maximum of probabilities.
\\Figure \ref{fig2} presents normalized probability
profiles for the analyzed knots.  It is visible that different knots show now quite similar type of profiles 
(e.g. knot $5_{1}$ and $5_{2}$) whereby the differences in the position of maximum between knots with different minimal number of
crossings can be easily perceived. It may be surprising that we observed here such a short optimal length for analyzed knots while earlier
studies showed that several hundred segments are needed to observe maximum occurrence of a given knot among closed walks of a given size
\cite{5,10,19}. This is simply due to the fact that our system takes into account the probability of cyclization.
\\In Figure \ref{fig3} we show the relation between the optimal length of random  knots and the length of the
corresponding ideal knots. This  relation is well approximated by a power law
function. Upon  fitting the free parameters of this function in the simulation data 
obtained for the knots with up to 7 crossings, we decided to check if by  knowing the length of ideal
configurations of more complicated knots we  can predict positions of the maximum of occurrence for the
corresponding  random knots. As the statistics of random knotting gets poor for knots  with increasing crossing
number we limited verifications of our  predictions to these knots with eight crossings which at their maxima of 
occurrence were represented more than 500 times out of 2$\cdot$10$^{9}$ random walks  with a given number of segments.
Analysis of our simulation data (Figure \ref{fig3})  positively verified our predictions for optimal sizes of random walks 
leading to the formation of these knots.
\\As already mentioned, ideal random chains have no thickness and this causes
that they reflect the behavior of polymers in the melt phase  where thin polymers have practically no exclusion
volume \cite{7,8}. However  when polymers are suspended in a good solvent, like DNA in aqueous  solution, the
exclusion volume of polymers becomes not negligible and  this strongly decreases the probability of forming
knots \cite{7}. It was observed  that the higher the effective diameter of the polymer the lower the  probability of
forming knots by random cyclization \cite{4,7,19}. We decided  therefore to investigate whether a power
law relation between the length  of ideal knots and the optimal length of randomly knotted chains also  holds for
chains with an exclusion volume. To this aim from our original  set of 2$\cdot$10$^{9}$ ideal random walks for every segment
length from 5 to 100 we  selected the walks which never showed a closer approach between any pair  of non
neighboring segments than the considered effective diameter (terminal segments of the chain are considered as neighboring
ones). Subsequently we  analyzed all configurations with approached ends for the types of formed  knots and
calculated the probabilities of various knots among all random  chains which fulfilled the criteria of a given
effective diameter. We observed that as the effective diameter grows the probability of forming various knots decreases and positions of
the maximum move toward longer chains. Figure \ref{fig3} (dashed line) shows the relation between  the length of ideal knots and the
optimal length of corresponding random knots formed by chains with the effective diameter being set to 0.05 of the segment length. The effective diameter
0.05 corresponds to this of  diluted solutions of DNA molecules in about 100 mM NaCl \cite{4}. In the case  of
DNA each segment in the random chain corresponds to 300 base pair long  region \cite{20}. It is visible that
the data can be again approximated by a power law function. Fact that lengths of ideal knots shows a 
correlation with the optimal sizes of corresponding random knots formed  by chains with a given effective diameter provides
another example that ideal knots are good predictors of physical behavior of real knots \cite{11}.
\begin{table}
\caption{Optimal sizes $O_{s}$ of random walks (in number of segments) leading to the formation of corresponding knots, the
length/diameter ratio $L_{D}$ values of ideal configurations of these knots $K_{n}$ \cite {23} and the values of the parameters in the fits of the
observed probabilities (see Fig. \ref{fig1}). The presented data are limited to knots with up to 7 crossings
since obtained by us, statistics for more complex knots is less good.}
\label{tableI}
\begin{tabular}{ccccccc}
\hline
$K_{n}$ & $O_{s}$ & $L_{D}$ & $a$ & $b$ & $d$ & $N_{0}$\\
\hline
$3_{1}$ & 25${\pm}$1 & 16.33 & (1.84${\pm}$0.01)${\times}$10$^{-1}$ & 1.57${\pm}$0.01 & 0.165${\pm}$0.001 & 5\\
$4_{1}$ & 42${\pm}$1 & 20.99 & (0.45${\pm}$0.01)${\times}$10$^{-1}$ & 2.24${\pm}$0.01 & 0.134${\pm}$0.001 & 6\\
$5_{1}$ & 54${\pm}$2 & 23.55 & (1.28${\pm}$0.02)${\times}$10$^{-2}$ & 2.65${\pm}$0.01 & 0.121${\pm}$0.001 & 7\\
$5_{2}$ & 56${\pm}$2 & 24.68 & (2.31${\pm}$0.04)${\times}$10$^{-2}$ & 2.77${\pm}$0.01 & 0.118${\pm}$0.001 & 7\\
$6_{1}$ & 74${\pm}$2 & 28.30 & (0.78${\pm}$0.03)${\times}$10$^{-2}$ & 3.75${\pm}$0.02 & 0.095${\pm}$0.001 & 7\\
$6_{2}$ & 75${\pm}$2 & 28.47 & (0.74${\pm}$0.03)${\times}$10$^{-2}$ & 3.67${\pm}$0.02 & 0.096${\pm}$0.001 & 7\\
$6_{3}$ & 76${\pm}$2 & 28.88 & (0.39${\pm}$0.02)${\times}$10$^{-2}$ & 3.69${\pm}$0.02 & 0.097${\pm}$0.001 & 7\\
$7_{1}$ & 89${\pm}$3 & 30.70 & (4.09${\pm}$0.47)${\times}$10$^{-7}$ & 3.95${\pm}$0.06 & 0.093${\pm}$0.001 & 8\\
$7_{2}$ & 92${\pm}$3 & 32.41 & (1.72${\pm}$0.16)${\times}$10$^{-3}$ & 4.33${\pm}$0.05 & 0.086${\pm}$0.001 & 8\\
$7_{3}$ & 92${\pm}$3 & 31.90 & (9.43${\pm}$0.85)${\times}$10$^{-4}$ & 4.03${\pm}$0.05 & 0.092${\pm}$0.001 & 8\\
$7_{4}$ & 97${\pm}$3 & 32.53 & (5.55${\pm}$0.67)${\times}$10$^{-4}$ & 4.25${\pm}$0.06 & 0.087${\pm}$0.001 & 8\\
$7_{5}$ & 97${\pm}$3 & 32.57 & (1.32${\pm}$0.09)${\times}$10$^{-3}$ & 4.24${\pm}$0.04 & 0.089${\pm}$0.001 & 8\\
$7_{6}$ & 98${\pm}$3 & 32.82 & (1.71${\pm}$0.14)${\times}$10$^{-3}$ & 4.36${\pm}$0.04 & 0.086${\pm}$0.001 & 8\\
$7_{7}$ & 95${\pm}$3 & 32.76 & (8.82${\pm}$0.96)${\times}$10$^{-4}$ & 4.31${\pm}$0.06 & 0.087${\pm}$0.001 & 8\\
\hline
\end{tabular}
\end{table}
\vspace{1.5mm}
\\Post factum it might seem to be obvious that knots requiring higher  length of the rope to tie them should require higher length
of a random  walk to reach their highest occurrence. However until recently the  minimal length of the rope to
tie a given knot was not known. In addition  the relation between the optimal length of random walk producing a
given  knot and the length of ideal knot was not yet proposed in the literature. On the other hand a simple
expectation would dictate that the shorter the length of ideal knot the higher the probability of its formation.
So for  example trivial knots are more frequent than trefoils and these are more frequent than $4_{1}$ knots. However
this does not hold for $5_{1}$ and $5_{2}$ knots.  Ideal knot $5_{1}$ is slightly shorter than ideal $5_{2}$ knot (which is
consistent  with the optimal size of random walks leading to the formation of corresponding knots), but $5_{2}$ knot
formation by random walks is circa  twice more frequent than formation of $5_{1}$ knot. Therefore the values of 
random knots probabilities (in contrast to the positions of the maxima) are not related by a relatively simple
growing function to the values of  lengths of the corresponding ideal knots.
\vspace{1.5mm}
\\What can be the possible
applications resulting from the determination  of the optimal size of knots? For chemical cyclization of polymer
chains  we can use linear polymer of a specific length and thus promote formation  of a given type of knot.
Materials with interesting properties could be  formed by this way.
\vspace{1.5mm}
\\{\bf Acknowledgment.}We thank Alexander
Vologodskii and Vsevolod Katritch for making available their routine for the calculation of Alexander polynomial. We also thank Piotr
Pieranski for numerous discussions. This work was supported 
by the Swiss National Science Foundation (31-58841.99 and 3152-061636.00).

\end{document}